\documentclass[12pt]{article}

\newcommand{\Si}{\Sigma}
\newcommand{\vp}{\varphi}
\newcommand{\ta}{\theta}
\newcommand{\vt}{\vartheta}
\newcommand{\K}{\cal K}
\newcommand{\al}{\alpha}

\newcommand{\wh}{\widehat}
\newcommand{\ep}{\epsilon}
\newcommand{\lb}{\lambda}
\newcommand{\La}{\Lambda}

\usepackage{amsmath}
\DeclareMathOperator{\tr}{tr}

\usepackage{graphicx,psfrag} 
\usepackage[utf8]{inputenc}   
\usepackage[T1]{fontenc}        
\usepackage[english]{babel}
\usepackage{amsmath,amssymb}
\usepackage{mathrsfs}
\usepackage{xcolor}

\usepackage{enumitem}
\usepackage{mathtools}
\usepackage[hidelinks]{hyperref}
\usepackage{amsthm}
\usepackage[numbers]{natbib}
\usepackage{float}
\usepackage{tikz-cd}

\usetikzlibrary{arrows.meta, patterns, calc}

\newcommand{\verticalEllipse}[4]{
    \pgfmathsetmacro{\centerY}{(#2 + #3)/2}
    \pgfmathsetmacro{\radiusY}{abs(#2 - #3)/2}
    \draw[ellipse style] (#1,\centerY) ellipse [x radius=#4 cm, y radius=\radiusY cm];
}

\newtheorem{theorem}{Theorem}

\newtheorem{proposition}{Proposition}
\newtheorem{prop}[proposition]{Proposition}
\newtheorem*{defn}{Definition}
\newtheorem{lem}{Lemma}

\theoremstyle{definition}

\newtheorem{example}{Example}
\newtheorem{remark}{Remark}
\allowdisplaybreaks
\numberwithin{equation}{section}

\usepackage[normalem]{ulem}

\title{Rigidity aspects of a cosmological singularity theorem }
\begin{document}

\maketitle

\begin{center}
{\large Eric Ling} \\
Copenhagen Center for Geometry and Topology (GeoTop),
Department of Mathematical Sciences,
University of Copenhagen\\
Universitetsparken 5, DK-2100 Copenhagen, Denmark
\\~\\ {\large Carl Rossdeutscher, Walter Simon and Roland Steinbauer}\\
Fakult\"at f\"ur Mathematik, Universit\"at Wien\\
Oskar-Morgenstern Platz 1, 1090 Wien, Austria
\end{center}

\begin{abstract}

Improving a singularity theorem in General Relativity  by Galloway and Ling  we show the following (cf.\ Theorem 1):
If a globally hyperbolic spacetime $M$ satisfying the null energy condition contains a closed, spacelike Cauchy surface $(V,g,\cal K)$ (with metric $g$ and extrinsic curvature ${\cal K}$)  which is  2-convex (meaning that the sum of the lowest two eigenvalues of ${\cal K}$ is non-negative), then either
$M$ is past null geodesically incomplete, or $V$  is  a spherical space, or $V$ or some finite cover $\tilde V$ is a surface bundle over the circle, with totally geodesic fibers.
Moreover,  (cf.\ Theorem 2) if $(V,g,{\cal K})$ admits a $U(1)$ isometry group with corresponding Killing vector $\xi$, we can relax the convexity requirement in terms of a decomposition of ${\cal K}$ with respect to the directions parallel and orthogonal to $\xi$.
Finally, (cf. Propositions 1-3) in the special cases that $V$ is either nonorientable, or non-prime, or an orientable Haken manifold with vanishing second homology, we obtain stronger statements in both Theorems without passing to covers. While our results do not use Einstein's equations, we provide 
several classes of $\Lambda$-vacuum solutions of these equations as examples.

\end{abstract}

\section {Introduction}

\label{int}

The classic singularity theorems due to Hawking and Penrose \cite{hawking1975large}  yield geodesic incompleteness in different settings, namely "cosmological" and "black hole" ones.
In the former setting one requires a compact Cauchy surface and a condition on its second fundamental form, while the latter setting assumes noncompact data containing a trapped or marginally trapped surface. In either case, convergence ("energy-") conditions on the spacetime Ricci tensor $\mathrm{Ric}$
are also necessary. In the sequel we will only need the null energy condition (NEC), namely $\mathrm{Ric}(X,X) \ge 0$ for all null vectors $X$.

 In the intervening decades, Hawking's and Penrose's results have been substantially modified and strengthened in various directions. In particular,
 Galloway and the first-named author obtained the following result. All manifolds and fields are 
 smooth ($C^{\infty}$) throughout the present paper.
 
 \begin{theorem}~ (Thms.\ 1 in \cite{Ling,GallowayLing25} and the Remark in Sec.\ 4 of \cite{Ling2})
\label{Gal_Ling} 
 \begin{enumerate}[label=(\arabic*)]

  \item Let $M$ be a (3+1) dimensional, globally hyperbolic spacetime which satisfies the NEC and has  a closed spacelike Cauchy surface $V$.
 
\item Assume $V$ is strictly $2$-convex, i.e. the sum of the smallest two eigenvalues of the future 
second fundamental form ${\K}$ is positive.

  \end{enumerate}
  Then at least one of the following scenarios applies:
  \begin{enumerate}
\item[(i)] $M$ is past null geodesically incomplete, or
\item[(ii)] $V$ is a spherical space.
  \end{enumerate}
\end{theorem}

We recall that a {\it spherical space} $V$ is by definition diffeomorphic to a quotient of the 3-sphere 
$\mathbb S^3$  , $V = \mathbb{S}^3/\Gamma$, where $\Gamma$
is isomorphic to a subgroup of $SO(4)$.
\begin{remark}
\label{rem0}
The above formulation indicates that conclusions $(i)$ and (ii) do not exclude each other. 
 A "truncated" de Sitter spacetime provides an example in which both conclusions hold.
 \end{remark}

Condition (2) on $\K$ in Theorem \ref{Gal_Ling} is rather restrictive and obviously 
excludes even the interesting time-symmetric case. The purpose of the present work is to relax this condition. 
To state our results we fix some notation, almost all of which coincides with \cite{Ling}.

\begin{figure}[h!]
\begin{tikzpicture}[scale=0.7]

  \coordinate (A) at (6,3);   
  \coordinate (B) at (18,3);  
  \coordinate (C) at (12,-2); 
  \coordinate (D) at (0,-2);  

  \draw[thick] (A) -- (D); 
  \draw[thick] (B) -- (C); 

  \draw[thick] (D) .. controls (5,0) and (9,-4) .. (C);

  \draw[thick] (A) .. controls (8,5) and (10,2) .. (B);

  \draw[very thick] (5,1.2) .. controls (7,1.8) and (9,1.6) .. (11,1.2) 
                    .. controls (12,0.8) and (12.2,0.2) .. (11.5,-0.2) 
                    .. controls (10.5,-0.6) and (9,-0.7) .. (7.5,-0.5) 
                    .. controls (6,-0.3) and (4.8,0.2) .. (4.5,0.6) 
                    .. controls (4.2,1) and (4.6,1) .. (5,1.2);
  
  \coordinate (P_sigma) at (11.8,.7); 
  \fill (P_sigma) circle (2pt);
  
  \coordinate (P_V) at (10,2.5);
  \fill (P_V) circle (2pt);
  
  \draw[-{Stealth[length=3mm]}, thick] (P_sigma) -- ++(2.8,1.5) node[below right] {$\nu$};
  
  
  \coordinate (end) at (13.8,-2.9);
  \coordinate (invisible) at (12.9,-1.3);
  \draw[thick, dashed] (P_sigma) -- (invisible);
  \draw[-{Stealth[length=3mm]}, thick] (invisible) -- (end);
  \node[right] at (end) {$\ell^+$};

  \coordinate (end) at (9.5,-3.5);
  \coordinate (invisible) at (10,-2.6);
  \draw[thick, dashed] (P_sigma) -- (invisible);
  \draw[-{Stealth[length=3mm]}, thick] (invisible) -- (end);
  \node[right] at (end) {$\ell^-$};
  
  \draw[-{Stealth[length=3mm]}, thick] (P_V) -- ++(.4,3.5) node[above right] {$u$};
  
  \node at (3.2,-.5) {\Large $V$};
  \node at (8,-1.2) {\Large $\Sigma$};

\end{tikzpicture}
\caption{The basic geometric setup}
\label{Fig:1}
\end{figure}
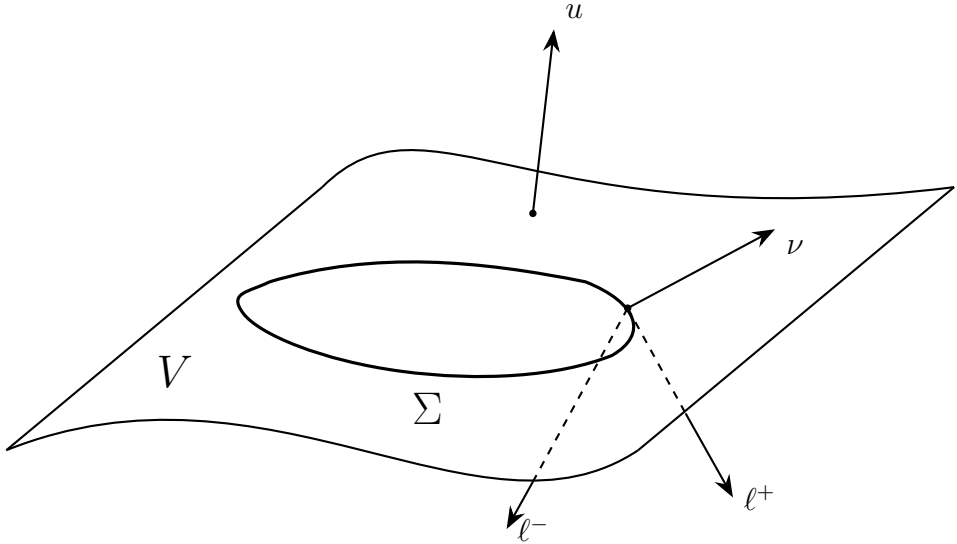

We denote the induced metric on $V$ by $g$ and the second fundamental form by $\K$ (in order to distinguish it from $K(\pi,1)$ manifolds). Let $u$ be the future directed timelike unit normal to $V$, and refer $\K$  to this direction. We next consider a two-sided embedding of a surface $\Sigma \subset V$. We choose some "outward"
 direction with unit normal $\nu$ on $\Sigma$ and denote by $H$ the corresponding
 outward mean curvature of $\Si$, cf.\ Figure \ref{Fig:1}. 

 The two past directed null normals to $\Si$ read $\ell^{\pm} = - u \pm \nu$,
 with corresponding null second fundamental forms 
 $\chi^{\pm} = {\cal P}_{\Si}(\nabla \ell^{\pm})$ (where $\nabla$ is the covariant derivative on $M$ 
and ${\cal P}_{\Si}$ the projection onto $\Si$)   
 and null expansions $\theta^{\pm} = \tr_{\Si}(\chi^{\pm})$ (where $\tr_{\Si}$ denotes the trace on $\Si$).   
 Marginally inner (outer) trapped surfaces (MITS, MOTS) are defined by $\theta^{-} = 0$ ~($\theta^{+} = 0$)  where 
 $\theta^{\pm}$ refer
 to the inward (outward) null directions $\ell^{\pm}$. 
We also recall the general decomposition
 \begin{equation}
 \label{dec}
 \theta^{\pm} = - \tr_{\Si} {\K} \pm H.
 \end{equation}


We can now state our main results.
\begin{theorem}\label{new} ~
\begin{enumerate}[label=(\arabic*)]
\item Let $M$ be a (3+1)-dimensional, globally hyperbolic spacetime which satisfies the NEC and has a closed spacelike Cauchy surface $V.$
\item Assume $V$ is $2$-convex, i.e.\ the sum of the smallest two eigenvalues of the future 
second fundamental form $\mathcal{K}$ is non-negative.
  
\end{enumerate} 
Then at least one of the following scenarios applies: 
    \begin{enumerate}[label=(\roman*)]
        \item $M$ is past null geodesically incomplete, or
        \item $V$ is a spherical space, or
        \item $V$ or a finite cover $\tilde{V}$ thereof is a surface bundle over the circle $\mathbb{S}^1$, where the fibers $\Sigma$ are totally geodesic within $V$ (or the finite cover $\tilde{V}$) and are MOTS/MITS within $M$ (or the corresponding spacetime cover $\tilde{M}$). 
    \end{enumerate}
\end{theorem}

\begin{remark} 
\label{rem1}
Extending Remark \ref{rem0}, 
 the "Misner-identified"  Taub-NUT space  (cf.\ Example \ref{ex2} in Section \ref{examples})  is inextensible as a globally hyperbolic spacetime. It  satisfies  requirements 
 $(1)$ and $(2)$, and both $(i)$ and (ii) apply. Moreover, spacetimes with data of type  
 (iii) can be geodesically complete or incomplete, as Examples \ref{ex3}, \ref{ex4}, \ref{ex6}, \ref{ex8} 
 show. 
  \end{remark}

There are important special cases in which we can strengthen the statements of Theorem \ref{new}. In the following three propositions, we obtain rigidity without having to pass to a cover. 
To state these results we recall that a \emph{Haken manifold} is a $P^2$-irreducible compact $3$-manifold which contains an embedded two-sided \emph{incompressible surface}. By the latter we mean
 a non-simply connected closed two-sided immersion $f\colon\Si \to V$ with $f_*\colon \pi_1(\Si) \rightarrow \pi_1(V)$ being injective.
Furthermore, by a \emph{semibundle} we mean $V$ is the union of two twisted $I$-bundles over a surface $S$. By fibers of a semibundle we refer to the fibers of $V \setminus (S \sqcup S)$, which is a surface bundle over an open interval with fibers $\Sigma$ which double cover $S$. 

\begin{prop}
\label{haken}
Under the assumptions of Theorem \ref{new}, if $V$ is an orientable Haken manifold with $H_2(V) = 0$, then at least one 
of the following scenarios applies:
\begin{enumerate}[label=(\roman*)]
    \item  $M$ is past null geodesically incomplete, or
    \item $V$ has semibundle structure, where the fibers $\Si$ are totally geodesic within $V$ and MOTS/MITS within $M$.
\end{enumerate}
\end{prop}

An example where (ii) in Proposition \ref{haken} holds is when $V$ is the flat Hantzsche-Wendt manifold (see Example \ref{ex8} in Section \ref{examples}). Then $V$ is orientable, Haken, and $H_2(V) = 0$. The Lorentzian product $M = \mathbb{R} \times V$ is complete and  $V$ satisfies conclusion (ii) of Proposition \ref{haken}. Note that this is a stronger statement than obtained from Theorem \ref{new}, since we do not have to go to a cover.
   
A similar statement can be made if $V$ is nonorientable. 

\begin{prop}
\label{orient}
Under the assumptions of Theorem \ref{new}, if $V$ is nonorientable, then at least one 
of the following scenarios applies:
\begin{enumerate}[label=(\roman*)]
     \item  $M$ is past null geodesically incomplete, or
    \item $V$ is diffeomorphic to a surface bundle over the circle $\mathbb{S}^1$, where the fibers $\Si$ are totally geodesic within $V$ and MOTS/MITS within $M$.
\end{enumerate}
\end{prop}

Below the proof of Proposition \ref{orient} in section \ref{proofs}, we give three examples where (ii) applies.

Recall that a three-manifold $V$ is \emph{prime} if it cannot be written as the connected sum of two three-manifolds $A$ and $B$ with neither of them being the three-sphere. 

\begin{prop}
\label{nonprime}
Under the assumptions of Theorem \ref{new}, if $V$ is non-prime, then at least one 
of the following scenarios applies: 
\begin{enumerate}[label=(\roman*)]
     \item  $M$ is past null geodesically incomplete, or  
    \item $V$ is diffeomorphic to $\mathbb{RP}^3\#\mathbb{RP}^3$, where the fibers $\Si\cong \mathbb{S}^2$ are totally geodesic within $V$ and MOTS/MITS within $M$.
\end{enumerate}
\end{prop}

An example where (ii) of Proposition \ref{nonprime} applies is the Nariai spacetime (see Example 4 in Section \ref{examples})  quotiented out by the double covering $\mathbb{S}^2 \times \mathbb{S}^1 \to \mathbb{RP}^3 \# \mathbb{RP}^3$ on the spatial slices.

Unfortunately,  the condition of $2$-convexity assumed above still excludes a lot of interesting cases. However,
in the presence of the cyclic isometry group $U(1)$ we can indeed weaken the signature requirements on $\K$ in terms of its decomposition with respect to the Killing vector $\xi$ corresponding to 
the $U(1)$ symmetry. Let $\|\xi\|$ be the norm of $\xi$,  
$\zeta = \|\xi\|^{-1} \xi$ the normalisation and $\zeta^*$ its dual.  We next define the projection 
${\cal P} = {\cal I} -  (\zeta \otimes \zeta^*)$ (a $(1,1)$- tensor) onto the 2-space orthogonal to $\xi$, 
 where ${\cal I}$ is the identity. Further, we denote by ${\K} (\xi,\xi)$ (a function) and ${\cal K}^{\bot}  = {\K(\bot,\bot)}
 = {\cal P}^T {\K}{\cal P} $   (a $(1,1)$-tensor), the projections of $\K$ parallel and orthogonal to  $\xi$. 
 
 In the sequel we will use the shorthand "data" for the set $(V,g,\K)$, still without assuming 
 any field equations. We obtain the following result.

 \begin{theorem}~ 
 \label{sym}
 \begin{enumerate}[label=(\arabic*)]
 \item Let $M$ be a (3+1)-dimensional globally hyperbolic spacetime which satisfies the NEC and has  a closed spacelike Cauchy surface $V$.
 Suppose further that the data $(V,g,\K)$ are invariant under the
  cyclic isometry group $U(1)$ with corresponding Killing vector $\xi$.
  \item Furthermore, suppose that 
\begin{equation}
\label{Kdec}
    {\K(\xi,\xi)} + \mu\|\xi\|^2 \ge 0 
\end{equation}
  
  where $\mu$ is the
  smaller eigenvalue of $\K^{\bot}$, i.e. $\mu \le \widehat \mu$ for the eigenvalues
  $\mu, \wh \mu$.
  \end{enumerate}
  
 Then at least one of the following scenarios applies:

    \begin{enumerate}[label=(\roman*)]
     \item  $M$ is past null geodesically incomplete, or
    \item $V$ is a spherical space, or
     \item $V$ or a finite cover $\tilde{V}$ thereof is a surface bundle over the circle $\mathbb{S}^1$, where the fibers $\Sigma$ are $U(1)$-symmetric and totally geodesic  with Euler number  $\chi(\Sigma) \geq 0$\footnote{Specifically, $\Sigma$ is either the torus or the 2-sphere if orientable or the Klein bottle or the two-dimensional projective space if nonorientable.} within $V$ (or the finite cover $\tilde{V}$). Moreover, theses fibers are MOTS/MITS within $M$ (or the corresponding spacetime cover $\tilde{M}$), or
    \item $V$ or a finite cover $\tilde{V}$ is diffeomorphic to a surface bundle over the circle $\mathbb{S}^1$ on which the given isometry acts. Moreover, the fibers $\Si$ are minimal  within $V$ or $\tilde V$.
    \end{enumerate}

 \end{theorem}
\begin{remark}
    \label{ev}
    In the presence of the $U(1)$ isometry, $2$-convexity indeed implies (1.2) and is thus more restrictive. This can be seen as follows: Assume condition (1) from the above theorem with normalized Killing vector $\zeta$ and that $V$ is $2$-convex. Let $\lambda_1 \le \lambda_2 \le \lambda_3$ be the eigenvalues of $\K$ and let $\mu \leq \widehat\mu$ be the eigenvalues of $\K^{\bot}$. 
    Then $\K(\zeta, \zeta) + \mu = \operatorname{tr}\K - \operatorname{tr}\K^\perp + \mu$ $= \lambda_{1} + \lambda_2 + \lambda_3 - \widehat{\mu} \geq \lambda_1 + \lambda_2 \geq 0$, where we used the Eigenvalue Interlace Theorem \cite{SF} to conclude that $\lambda_3 - \widehat{\mu} \geq 0$. Therefore, condition (1.2) is satisfied.
    
Examples reveal that condition (2) of  Theorem \ref{sym} is in fact significantly less restrictive than $2$-convexity.
In particular, while $2$-convexity excludes the family of so-called ($t$-$\phi$)-symmetric data,
the latter are all admitted by (\ref{Kdec}) (cf.\ Section \ref{examples} for details).
    
\end{remark}
\begin{remark}
\label{amb}
Apart from the ambiguities mentioned in Remarks \ref{rem0} and \ref{rem1} we mention the following: 
It is clear that (iii) and (iv) are exclusive with respect to a given Killing field $\xi$, since 
it cannot be both tangent to the base and to the fibers. However, 
when the data $(V,g,\K)$ enjoy a symmetry group containing $U(1) \times U(1)$, (iii) and (iv) can both apply with respect to the different factors; we refer to Examples \ref{ex3} and \ref{ex6} in Section \ref{examples}.
\end{remark}

\begin{remark}
There are Propositions \ref{haken}, \ref{orient} and \ref{nonprime}
in the setting of Theorem \ref{sym} {\it mutatis mutandis}, more precisely:
\begin{itemize}
    \item Points (ii) in the propositions could be reformulated as in  (iii) of Theorem \ref{sym}.
    
    \item A further alternative  (iii) has to be added in each proposition corresponding to point (iv)
    of Theorem \ref{sym}. 
\end{itemize}

\end{remark}

This paper is organized as follows. The subsequent Section \ref{pre} contains
 three auxiliary results, while the cores of the proofs of the results stated above are given in Section \ref{proofs}. As mentioned earlier the final Section \ref{examples} consists of examples. 
\medskip

We conclude here with sketching the strategy of our proofs.
For simplification, we assume that $V$ is not a spherical space, and we  pass to a finite, orientable cover $\tilde V$ whenever necessary or useful.
Our key prerequisite is the  prime decomposition  of orientable 3-manifolds, see (\ref{prime}) below. 
Moreover, from  the positive resolution of the virtual positive $b_1$-conjecture \cite{Agol}, it follows that $V$ or a finite cover $\tilde{V}$
has non-vanishing second homology $H_2$. As is well-known, this implies that $V$ (or $\tilde V$) contains an embedded, non-separating, two-sided minimal surface $\Si$ of least area.
By a known Lemma (Lemma 0), it follows that a neighborhood of $\Si$ can be foliated by CMC 
surfaces $\Si_t$.  To prove Theorem 1 we now invoke  the convexity requirement on ${\cal K}$ which, via (\ref{dec}), yields control on the signs of the  null expansions $\ta^{\pm}$ on each $\Si_t$. Together with the null energy condition,  this allows an application  of Penrose's singularity theorem on a suitable infinite cover of $V$ (or $\tilde V$). This gives past null geodesic incompleteness except for the degenerate cases stated in (ii), (iii) of Theorem \ref{new}. In case (iii) compactness theorems allow us to extend the local folitaion of Lemma 1 to all of $V$.

The proof of Theorem 2 deviates from above as follows: Setting out from the local CMC 
foliation provided by Lemma 0, Lemma 2 shows that the given isometry is either everywhere tangent to all leaves, or nowhere tangent to any leaf. 
In the former case, the less restrictive convexity condition (\ref{Kdec}) now takes the role of $2$-convexity and yields analogous results as before,
while in the latter case we end up with  the additional alternative (iv) of Theorem \ref{sym}.

We choose not to sketch the proofs of Propositions 1-3 since they are highly technical.

\begin{remark} 
\label{Greg}
{We remark that Theorem \ref{new} can be proven in a different way, namely by using the positive resolution of the virtual Haken conjecture and Proposition \ref{haken}.  In fact, this detour was taken by the authors originally. We are grateful to Greg Galloway for bringing to our attention the positive resolution of the virtual positive $b_1$ conjecture which greatly simplifies the proof. A similar simplification also applies to the proof of the main result in \cite{Ling}, reformulated as Theorem 0 above (see \cite{GallowayLing25}).}
\end{remark}

Apart from our main reference \cite{Ling}, the recent paper \cite{Ling2} provided some inspiration for the present work as well. In particular there is some overlap between Theorems 1 and 2 above and Theorem 9 in \cite{Ling2}. While the latter amounts to an extension  of Theorem 0 in the case 
$H_2(V) \neq 0$, in this paper we also address the more subtle case $H_2(V) = 0$.

\section{Preliminaries}

\label{pre}

Here we collect some preliminary results.

To begin with, recall that for a minimal surface $\Si$ the \emph{stability operator} $\mathcal{L}(\psi)$ for $\psi \in C^\infty(\Sigma)$ is given by 
\begin{equation}\label{sop}
\delta_{\psi \nu} H =  {\cal L}_\nu(\psi) = - \Delta_{\Si} \psi -
\frac{\psi}{2} \left (R_V - R_{\Si} + {\cal C}_{\Si} (t \otimes t) \right).
\end{equation}
 In (\ref{sop})  $R_V$ and $ R_{\Si}$ are the Ricci scalars  of the manifolds in the subscripts,
$t$ is the trace-free part of the extrinsic curvature and ${\cal C}_{\Si}$ is a contraction (over both indices).

A minimal surface $\Si$ is called \emph{stable (strictly stable, marginally stable)} if the lowest eigenvalue $\lb$  of ${\cal L}(\psi)$
satisfies $\lb \ge 0$ ($\lb > 0$, or $\lb = 0$, respectively).

The following lemma is a standard application of the inverse function theorem. For a proof of the more involved $\lambda = 0$ case, see the proof of Thm.\ 2.38 in \cite{DL} or \cite[Appendix A]{Ling2}.

\begin{lem} 
Let $\Sigma$ be an embedded stable minimal surface within a Riemannian mani\-fold $V$. Then there is a neighborhood $U$ of $\Sigma$ within $V$, diffeomorphic to $(-\ep, \ep) \times~\Sigma$, such that  each leaf $\Sigma_t := \{t\} \times \Sigma$ has constant mean curvature.
\label{cmc}
\end{lem}

\begin{lem}
\label{pen} 
Under the assumptions of Theorem \ref{new}, assume further that 
\begin{enumerate}[label=(\arabic*)]
\item $M$ is past null geodesically complete,
\item $V$ contains an embedded, two-sided,   minimal surface $\Sigma$,
\item There exists a noncompact cover $p\colon \tilde{V} \to V$  and a lift $\tilde{\Sigma}$ of $\Sigma$ which separates $\tilde{V}$ into two disjoint noncompact open submanifolds.
\end{enumerate}
Then there exists a neighborhood $U$ of $\Sigma$ in $V$ which is diffeomorphic to $(-\epsilon, \epsilon)\times \Sigma$ and each $\Sigma_t = \{t\} \times \Sigma$ is totally geodesic within $V$.
\end{lem}

\noindent {\bf Proof.} The following facts will be used to prove this Lemma as well as Lemma \ref{Kil}. First, there is a spacetime cover $P \colon \tilde{M} \to M$ such that $p = P|_{\tilde{V}}$, $\tilde{V}$ is a Cauchy surface for $\tilde{M}$, and the second fundamental form on $\tilde{V}$ is the pullback $\tilde{\K} = P^* \K$, see \cite[Lem. 4]{Ling}. Moreover, the null convergence condition also ``lifts'', and $M$ is null geodesically complete if and only if $\tilde{M}$ is. Lastly, if a surface $\Sigma'$ embedded in $V$ is constructed via a foliation from $\Sigma$ within $V$,  then $\Sigma'$ lifts to some $\tilde{\Sigma}'$, which, like $\tilde{\Sigma}$, also separates $\tilde{V}$ into two noncompact ends. 

 Let $\lambda$ denote the principal eigenvalue for the stability operator ${\cal L}$ of $\Sigma$ within $V$. We show that $\lambda = 0$. Let $\phi > 0$ be the corresponding principal eigenfunction. Consider the variation $\Sigma_t$ given by $x \mapsto \exp_x\big(t\phi(x)\nu(x) \big)$ for $x \in \Sigma$. Then
 \[
 \frac{\partial H}{\partial t}\bigg|_{t = 0} =  {\cal L}(\phi) = \lambda \phi.
 \] 
If $\lambda < 0$, then $H(t) < 0$ for small $t > 0$. By the $2$-convexity of $\mathcal{K}$, it follows that $\theta^+(t) < 0$ for these small $t$, cf.\ \eqref{dec}. Therefore an application of Penrose's singularity theorem (Thm.\ 7.1 of \cite{Andersson_2009}) applied to the outer trapped surface $\tilde{\Sigma}_t$ within $\tilde{M}$ implies past null geodesic incompleteness within $\tilde{M}$ and hence also in $M$, which is a contradiction. Similarly, if $\lambda > 0$, then $H(t) < 0$ for some values $t < 0$. Again, an application of Penrose's singularity theorem in the cover yields a contradiction. Thus $\lambda = 0$. 

By Lemma \ref{cmc} there is a neighborhood $U$ diffeomorphic to $(-\epsilon, \epsilon) \times \Sigma$ such that each leaf $\Sigma_t$ has constant mean curvature. We distinguish two cases.

\begin{description}
 \item[1.)  $\mathbf{H_t \neq 0 \text{ for some }t.}$] If $H_t > 0$, then $\theta^-_t < 0$. If $H_t < 0$, then $\theta^+_t < 0$. In either case, we can apply Penrose's singularity theorem in the cover since both components of $\tilde{V} \setminus \tilde{\Sigma}_t$ are noncompact, a contradiction.

 \item[2.)  $\mathbf{H_t \equiv 0 ~\forall~ t}$.] We distinguish two more cases.

\begin{description}

            \item [a.) $tr_{\Si_t} {\K} \not\equiv 0$.] 
            By the $2$-convexity of $\K$, both null expansions satisfy $\theta^{\pm}_t \le 0$ and hence $\tilde{\theta}^{\pm}_t \le 0$ but are not identically vanishing. Hence for either choice, Lemma 5.2 of \cite{AM} yields a deformation  $\tilde{\Si}'$ of $\tilde{\Si}_t$  such that (at least) the chosen $\tilde{\theta}'$ satisfies $\tilde{\theta}' < 0$ on $\tilde{\Si}'$. As above Penrose's theorem in $\tilde{M}$ contradicts past null geodesic completeness within $M$. 
             \item [b.) $tr_{\Si_t} {\K} \equiv 0$.]
            Now obviously $\tilde{\theta}_t^{\pm} \equiv 0$ for both signs. In terms of the null second fundamental forms $\tilde{\chi}_t^{\pm}$ defined in the Introduction, we first assume that 
            at least one of 
             \begin{equation}
            \label{W}
            \tilde{W}_t^{\pm} = |\tilde{\chi}_t^{\pm}|^2 + \tilde{\mathrm{Ric}}(\tilde{\ell}_t^{\pm},\tilde{\ell}_t^{\pm}),
            \end{equation}
              say $\tilde{W}_t^{+}$, does not vanish identically on $\tilde{\Si}_t$.  By virtue of Raychaudhuri's equation,
              $
            \frac{d}{ds}\tilde{\theta}_t^{+} =   - \tilde{W}_t^{+},
              $
              and another application of Lemma 5.2 in \cite{AM} we find surfaces $\tilde{\Sigma}'$ near the past light cone of $\tilde{\Sigma}_t$ with $\tilde{\theta}^{\prime +} < 0$. 
              Another application of Penrose's theorem in the cover $\tilde{M}$ and projecting down to $M$ contradicts past null geodesic completeness within $M$. Therefore both $(+)$-terms on the r.h.s.\ of (\ref{W}) have to vanish individually. Clearly, the case $\tilde{W}^-_t$ proceeds analogously. Thus the family $\Sigma_t$ is totally geodesic within $V$ since its second fundamental form is given by $\chi^+_t - \chi^-_t$. 
             \end{description}
\end{description}
$\hfill \Box$
\begin{lem}  
\label{Kil}
Under the assumptions of Theorem \ref{sym}, assume  that 
\begin{enumerate}[label=(\arabic*)]
\item $M$ is past null geodesically complete,
\item $V$ contains an embedded, two-sided,  minimal surface $\Sigma$ which is least area in its homology class.
\item There exists a noncompact cover $p\colon \tilde{V} \to V$ and a lift $\tilde{\Sigma}$ of $\Sigma$ which separates $\tilde{V}$ into two disjoint noncompact open submanifolds.
\end{enumerate}
Then either (i) or (ii) holds.
\begin{enumerate}[label=(\roman*)]
\item There exists a neighborhood $U$ of $\Sigma$ in $V$ which is diffeomorphic to $(-\epsilon, \epsilon)\times \Sigma$ and each $\Sigma_t = \{t\} \times \Sigma$ is totally geodesic with $\xi$ tangent to it. Moreover, $\Si$ or its orientable double cover $\tilde \Si$ has spherical or toroidal topology.  
\item $V$ is diffeomorphic to a surface bundle over the circle $\mathbb{S}^1$ on which the given isometry acts; the fibers are minimal. 
 \end{enumerate}
\end{lem}

\noindent {\bf Proof}.  $\Si$ being least area implies non-negativity of the second variation of area
and therefore stability. Applying Lemma \ref{cmc} 
provides a smooth CMC foliation $\Si_t$. All subsequent statements apply to  sufficiently small one-sided 
 neighborhoods of $\Si = \Si_0$; w.l.o.g. we choose $t \geq 0$, and we take the ``outward'' direction  -- in particular the normal $\nu$ to the leaves -- pointing towards increasing $t$.

As in  Lemma \ref{pen}  we distinguish two main cases,  but now subtleties in point 1. require  due attention.
 
\begin{description}
  \item[1.)  $\mathbf{H_t \neq 0 ~for~ some ~ t \neq 0}$.]
We observe that by a mean value argument there is a neighborhood of some $s$ such that for all $\Si_t$ in that neighborhood $H_t > 0$
and strictly monotonically increasing in the outward direction. 


The  next step is to decompose the given Killing vector $\xi$ into the normal and tangential parts to $\Si_s$, viz. $\xi =\xi^{\bot}  + \xi^{\parallel} = \psi \nu + \xi^{\parallel}$ where $s$ is the point selected above and $\psi \in C^{\infty}$ is the  "lapse" of the foliation. 
Now the key step is to  show that  $\xi$ is tangent to all $\Si_t$ in the neighborhood of $s$
i.e. $\xi^{\bot} = \psi \nu \equiv 0$. ~

We first observe that this tangency necessarily holds on a strictly stable $\Si$  as the following calculation (cf.\ e.g.\ Prop.\ 2.12 in \cite{DL}) shows 
\begin{equation}
\label{delH}
0 = \delta_{\xi} H = \delta_{\psi \nu} H  = {\cal L}_{\nu} \psi.
\end{equation}
Here we have used that the variation of $H$ along the Killing vector $\xi$ vanishes on $\Si$, and definition (\ref{sop}) of the stability operator ${\cal L}_{\nu}$.
But if $\psi \not\equiv 0$, (\ref{delH}) implies that $\psi$ is an eigenfunction of ${\cal L}_{\nu}$ with eigenvalue zero,
which contradicts strict stability.

In the general case where $\Si$ is just least area rather than strictly stable we proceed as follows. We  assume  again $\psi \not\equiv 0$ and arrive at a contradiction.

Choosing a CMC surface $\Si_s$ as above, namely with $H_s > 0$ and strictly monotonically increasing outward, the flow corresponding to the isometry generates a family $(\Sigma_{s, u})_{u \in (-\ep, \ep)}$ of surfaces around $\Sigma_{s, 0} = \Si_s$ which all have the same mean curvature $H_s$ as $\Si_s$, see Figures 2 and 3.

\begin{figure}[h!]
\centering
\begin{tikzpicture}[scale=1.2,
    thick line/.style={line width=1pt},
    normal line/.style={line width=0.5pt},
    ellipse style/.style={draw, thick line},
    arrow style/.style={->, >=latex, line width=1pt},
    dashed style/.style={dashed, dash pattern=on 4pt off 4pt}
]

\draw[thick line]
    (4.59,-1.305)
    .. controls (5.5,-1.48) and (6.2,-1.75) .. (6.9,-2.05)
    .. controls (7.6,-2.35) and (8.3,-2.62) .. (9.0,-2.85)
    .. controls (9.7,-2.98) and (10.4,-3.05) .. (11.1,-3.05)
    .. controls (11.8,-3.02) and (12.5,-2.85) .. (13.275,-2.25);

\draw[thick line]
    (4.68,-7.92)
    .. controls (5.5,-7.78) and (6.2,-7.52) .. (6.9,-7.20)
    .. controls (7.6,-6.88) and (8.3,-6.58) .. (9.0,-6.35)
    .. controls (9.7,-6.20) and (10.4,-6.15) .. (11.1,-6.18)
    .. controls (11.8,-6.25) and (12.5,-6.50) .. (13.185,-6.975);

\verticalEllipse{5.13}{-1.42}{-7.85}{0.5}

\verticalEllipse{6.705}{-1.98}{-7.29}{0.45}

\verticalEllipse{8.595}{-2.72}{-6.51}{0.4}

\begin{scope}
    \pgfmathsetmacro{\centerY}{(-3.03 + -6.15)/2}
    \pgfmathsetmacro{\radiusY}{abs(-3.03 - (-6.3))/2}
    \draw[ellipse style] (9.9,\centerY) ellipse [x radius=0.35cm, y radius=\radiusY cm];
\end{scope}

\begin{scope}[shift={(6.8625,-5.285)}, rotate=-35]
    \draw[ellipse style, dashed] (0,0) ellipse [x radius=0.6cm, y radius=3.08cm];
\end{scope}

\draw[arrow style] (6.705,-4.7) -- (3.915,-4.8);
\draw[arrow style] (6.705,-4.7) -- (8.5,-7.5);


\node[anchor=west] at (8.3,-5.4) {$\Sigma_y$};
\node[anchor=west] at (4.995,-1.1) {$\Sigma_x$};
\node[anchor=west] at (9.65,-5.4) {$\Sigma$};
\node[anchor=west] at (6.48,-1.7) {$\Sigma_s$};
\node[anchor=west] at (3.8,-5.15) {$\nu$};
\node[anchor=west] at (7.05,-3) {$\Sigma_{s,u}$};
\node[anchor=west] at (7.7,-7.5) {$\ell^-$};

\fill (5.15,-7.82) circle (2pt);
\node[anchor=west] at (4.85,-8.2) {$x$};

\fill (8.65,-2.75) circle (2pt);
\node[anchor=west] at (8.5,-2.4) {$y$};

\end{tikzpicture}
\caption{The setup in case 1.)}
\end{figure}

\begin{figure}[h!]
 \centering
\begin{tikzpicture}[scale=1.5]
\tikzset{
  myarrow/.style={
    ->,
    >=stealth,
    line width=0.6pt,
    shorten >=1pt,
    shorten <=1pt
  }
}

\coordinate (center) at (3.0,-2.0);

\coordinate (x_point) at (0.27,-1.2);  
\coordinate (y_point) at (4.5,-1.9);  

\draw[thick] (0.5,-1.0) .. controls (-0.2,-1.5) and (-0.2,-2.5) .. (0.5,-3.0)
             .. controls (1.5,-4.0) and (2.5,-4.5) .. (3.0,-4.5)
             .. controls (3.5,-4.5) and (4.5,-4.0) .. (5.5,-3.0)
             .. controls (6.2,-2.5) and (6.2,-1.5) .. (5.5,-1.0)
             .. controls (4.5,-0.2) and (3.5,-0.2) .. (3.0,-0.2)
             .. controls (2.5,-0.2) and (1.5,-0.2) .. (0.5,-1.0);

\begin{scope}[shift={(center)}, scale=0.75, shift={(-3.0,2.0)}]
\draw[thick] (0.5,-1.0) .. controls (-0.2,-1.5) and (-0.2,-2.5) .. (0.5,-3.0)
             .. controls (1.5,-4.0) and (2.5,-4.5) .. (3.0,-4.5)
             .. controls (3.5,-4.5) and (4.5,-4.0) .. (5.5,-3.0)
             .. controls (6.2,-2.5) and (6.2,-1.5) .. (5.5,-1.0)
             .. controls (4.5,-0.2) and (3.5,-0.2) .. (3.0,-0.2)
             .. controls (2.5,-0.2) and (1.5,-0.2) .. (0.5,-1.0);
\end{scope}

\begin{scope}[shift={(center)}, scale=0.5, shift={(-3.0,2.0)}]
\draw[thick] (0.5,-1.0) .. controls (-0.2,-1.5) and (-0.2,-2.5) .. (0.5,-3.0)
             .. controls (1.5,-4.0) and (2.5,-4.5) .. (3.0,-4.5)
             .. controls (3.5,-4.5) and (4.5,-4.0) .. (5.5,-3.0)
             .. controls (6.2,-2.5) and (6.2,-1.5) .. (5.5,-1.0)
             .. controls (4.5,-0.2) and (3.5,-0.2) .. (3.0,-0.2)
             .. controls (2.5,-0.2) and (1.5,-0.2) .. (0.5,-1.0);
\end{scope}

\begin{scope}[shift={(center)}, scale=0.3, shift={(-3.0,2.0)}]
\draw[thick] (0.5,-1.0) .. controls (-0.2,-1.5) and (-0.2,-2.5) .. (0.5,-3.0)
             .. controls (1.5,-4.0) and (2.5,-4.5) .. (3.0,-4.5)
             .. controls (3.5,-4.5) and (4.5,-4.0) .. (5.5,-3.0)
             .. controls (6.2,-2.5) and (6.2,-1.5) .. (5.5,-1.0)
             .. controls (4.5,-0.2) and (3.5,-0.2) .. (3.0,-0.2)
             .. controls (2.5,-0.2) and (1.5,-0.2) .. (0.5,-1.0);
\end{scope}

\draw[thick, dashed] (x_point) .. controls (0.2,-1.4) and (.2,-1.7) .. (1.2,-2.5)
                     .. controls (2.5,-3.5) and (3.5,-3.8) .. (4.0,-3.2)
                     .. controls (4.5,-2.8) and (4.6,-1.8) .. (y_point)
                     .. controls (4.3,-.7) and (3.0,-0.8) .. (2.0,-0.8)
                     .. controls (1.0,-0.8) and (0.8,-0.9) .. (x_point);

\fill (x_point) circle (1pt);
\fill (y_point) circle (1pt);

\draw[myarrow] (1.35,-2.95) -- (1.1,-2.6); 
\node[below left] at (1.3,-2.95) {$\xi$};
\draw[myarrow] (4.5,-.96) -- (4.5,-1.35); 
\node[above right] at  (4.5,-.96) {$\xi$};
\draw[myarrow] (4.85,-2.8) -- (4.5,-2.7); 
\node[below right] at  (4.85,-2.8) {$\xi$};

\draw[myarrow] (1.2,-1.22) -- (1.1,-0.9); 
\node[below right] at (1.2,-1.2) {$\xi$};
\draw[myarrow] (.8,-1.7) -- (.55,-1.3); 
\node[below right] at(.8,-1.7) {$\xi$};

\node at (5.65,-2.5) {$\Sigma_x$};
\node at (5.05,-2.1) {$\Sigma_s$};
\node at (3.7,-2) {$\Sigma$};
\node at (.3,-2) {$\Sigma_{s,u}$};
\node at (4.15,-2.3) {$\Sigma_y$};
\node[above left] at (x_point) {$x$};
\node[above right] at (y_point) {$y$};

\end{tikzpicture}
\caption{The setup in case 1.)}
\end{figure}

Now each $\Sigma_{s, u}$ intersects the leaves of the given CMC foliation in a neighborhood of $\Si_s$ such that for $\lvert u \rvert$, $\lvert v \rvert$ small enough, $\Sigma_{s,u}$  touches some $\Si_x$ from the 
inside at some point $x$ such that $0 < H_s < H_x$, while $\Sigma_{s,v}$ touches some $\Si_y$ from the outside at some point $y$ with $ 0 < H_y < H_s$. 
We note that the points $x$ and $y$ always lie to the exterior and interior of $\Si_s$, respectively. This fact  entails that, if $\xi^{\bot}$  changes direction on $\Si_s$ 
both points lie on the same surface $\Sigma_{s,u} = \Sigma_{s,v}$ for some $u = v$; 
this case is depicted  in Figures 2 and 3. 

In any case, and at either point this contradicts the  maximum principle for the quasilinear mean curvature operators acting on the graph functions of the respective pairs of surfaces near their touching points (cf.\ Thm.\ 2.4. of \cite{AGH}) unless the 
family  $(\Sigma_{s, u})$ agrees with $\Si_s$, a contradiction. In particular $\psi \equiv 0$  as claimed. 

Finally, we are  in the position to invoke condition (2) of Theorem \ref{sym}. Using  the notation introduced in Section \ref{int} we write   $\zeta = \xi/\|\xi\|$ for the normalized Killing vector and ${\cal K}^{\bot} = {\cal K}(\bot,\bot) = {\cal P}^T{\cal K}{\cal P}$ where ${\cal P} $ is the projection orthogonal to $\xi$.

Let ${\cal O}$ be the orthogonal matrix which diagonalizes ${\cal K}^{\bot}$, i.e.
${\bar {\cal K}^{\bot}} = {\cal O} {\cal K}^{\bot} {\cal O}^{T} =
\mathrm{diag}(\mu, \widehat \mu)$, where we choose $\mu \le \widehat \mu$. 
We also introduce an orthonormal basis ($\zeta, \rho$) of the tangent space to $\Si_s$,
 define $\bar \rho = {\cal O} \rho$
and denote the components of $\bar \rho$ by $\bar \rho_1, \bar\rho_2$.

We obtain
\begin{eqnarray}
\label{trK}
\|\xi\|^2 \tr_{\Si_{s}} {\cal K} & = &  \|\xi\|^2 \left({\cal K}(\zeta,\zeta) + {\cal K}^{\bot}(\rho,\rho)\right)
=  {\cal K}(\xi,\xi) + \| \xi\|^2  \bar {\cal K}^{\bot} (\bar \rho, \bar \rho)  = \nonumber  \\
& = & {\cal K}(\xi,\xi) +  \|\xi\|^2 ( \mu \bar \rho_1^2 +     \widehat \mu \bar \rho_2^2 )
\ge  {\cal K}(\xi,\xi)  + \|\xi\|^2 \mu
\end{eqnarray}
since $\bar \rho$ is also a unit vector.

Therefore, from condition (\ref{Kdec}) of Theorem \ref{sym}, $\tr_{\Si_{s}} {\cal K} \ge 0$, and thus (\ref{dec})  implies that $\theta^-_s < 0$. As in the proof of Lemma \ref{pen} this property is inherited by the cover, i.e. 
$\tilde{\theta}^-_s < 0$, and application of Penrose's singularity theorem (with respect to the noncompact end in the $-\nu$\,-direction from $\tilde \Si$) contradicts the required null geodesic completeness within $M$.

  \item[\bf 2.) $\mathbf{H_t \equiv 0 ~\forall~ t}$.]
Now $\Si$ is marginally stable and each $\Sigma_t$ is minimal.  There are two more cases to distinguish.

\begin{description}
 \item[a.) For all leaves $\Si_t$,  $\xi^{\bot} = \psi \nu \equiv 0$]. ~ \\
 Using calculation (\ref{trK}) and condition (\ref{Kdec}) of Theorem 2, we now conclude 
  $\tr_{\Si_{t}} {\cal K} \ge 0$, and thus (\ref{dec})  implies that for every minimal surface $\Si_t$,
   we have $\theta^{\pm} \le 0$.


From here on, we can continue with the reasoning of the proof of point 2 of Lemma \ref{pen}. This yields  conclusion (i) except for the statement on the topology of $\Si$.

To show the latter we recall the relations  $\chi = 2(1-\mathfrak g)$  and $\chi = 2-\mathfrak g$ between  genus $\mathfrak g$ and Euler number $\chi$ for orientable and nonorientable 2-surfaces, respectively.
The Euler number is also the sum of all indices of the (isolated) zeros of $\xi$, by virtue
of the Poincar\'e-Hopf theorem.   The zeros of $\xi$ are necessarily isolated since $\Sigma$ is two-dimensional \cite[Thm. 8.1.5]{petersen2016riemannian}, and the index of an isolated zero of a Killing vector field is +1 since they're rotations in a neighborhood of the zero. 
 This restricts the topology of an orientable surface $\Si$ (or of its orientable double cover 
 $\tilde \Si$) to the sphere or the torus, and to the projective plane or the Klein bottle for
 a nonorientable $\Si$.

\item [b.) There exists a leaf $\Si$ such that $\xi^{\bot} = \psi \nu \not \equiv 0$].

Now calculation (\ref{delH}) does imply that the lapse  $\psi$  is an eigenfunction of ${\cal L}_{\nu}$ with eigenvalue zero.
By stability, $\psi$ is the unique (up to a constant) principal eigenfunction of ${\cal L}$ which has constant sign. Reversing the direction of $\xi$  if convenient yields
$\psi > 0$. Since the given isometry forms a $U(1)$ group this implies conclusion (ii), i.e. a foliation  by disjoint leaves $\Si_t$ which are all isometric to each other. 
Note that requirement (2) of Theorem \ref{sym}  is not used in this case. Accordingly, we do not obtain totally geodesic leaves in general.

\end{description}
\end{description}

$\hfill \Box$

\begin{remark} The fixed point set of an isometry on a  $3$ dimensional manifold is either empty or a
(possibly disconnected) one-dimensional subset, cf.\ \cite{SK},  Thm. II.5.3.(1).
In the preceding Lemma, the case of the free action  corresponds to the toroidal case of (i) 
or to case (ii), while the one-dimensional set of fixed points are the circles formed by 
the axes points of the leaves  in case (i). See Examples \ref{ex3}--\ref{ex8} in Section \ref{examples}.
\end{remark} 

\section{Proofs of the main results}

\label{proofs}

In connection with the subsequent proof, we recall the \emph{prime decomposition} of closed $3$-manifolds: each such orientable $3$-manifold $V$ can be decomposed uniquely into a finite connected sum of primes $V_i$, viz.\ \cite{Hatcher}
\begin{equation}
\label{prime}
V = V_1\, \# \, V_2 \, \#\dotsb \# \, V_k,
\end{equation}
where each prime is either a spherical space, diffeomorphic to $\mathbb{S}^2 \times \mathbb{S}^1$, or a $K(\pi,1)$-manifold. \\

\noindent {\bf Proof of Theorems \ref{new} and \ref{sym}.} 
 Assume that $V$ is not a spherical space and $M$ is past null geodesically complete.  We can assume $V$ is orientable by passing to the orientable double cover if necessary. 

We claim that $V$ or a finite cover thereof has non-trivial second homology. To see this, recognize that by the prime decomposition of three-manifolds, $V$ is homeomorphic to $V_1 \# \dotsb \# V_k$, where each $V_i$ is a prime manifold. Consequently,  $H_2(V) = H_2(V_1) \oplus \dotsb \oplus H_2(V_k)$. If some $V_i$ is topologically $\mathbb{S}^1 \times \mathbb{S}^2$, then $H_2(V) \neq 0$. If some $V_i$ is aspherical (i.e., the universal cover of $V_i$ is contractible), then $V_i$ has a finite cover with nonvanishing second homology by the positive resolution of the virtual $b_1 > 0$ conjecture \cite{Agol}; in this case,  $V$ also has a finite cover with non-trivial second homology (see \cite[Lem. 6]{Ling}). Lastly, if each $V_i$ is a spherical space, then the arguments below the proof of \cite[Lem. 6]{Ling} show the desired result. This proves the claim.

Thus we can assume $H_2(V) \neq 0$. From well-known results in geometric measure theory, $V$ contains an embedded, nonseparating, two-sided, minimal, least-area surface $\Sigma$.

We first prove Theorem \ref{new} by showing that conclusion (iii) holds. Since $\Sigma$ is two-sided and nonseparating, we can cut $V$ along $\Sigma$ to produce a compact manifold $W$ with two boundary components -- each one isometric to $\Sigma$. We construct a covering $\tilde{V}$ of $V$ by gluing $\mathbb{Z}$ copies of this compact manifold $W$ end to end.  See e.g.\ the proof of Prop.\ 5 of \cite{Ling}.  It's clear that the hypotheses of Lemma \ref{pen} are satisfied for some particular copy of $\Sigma$ in the cover $\tilde{V}$. Thus there is a neighborhood of $\Sigma$ within $V$, diffeomorphic to $(-\epsilon, \epsilon) \times \Sigma$, such that its leaves $\Sigma_t$ are totally geodesic for all $t$. 
Using the  traced Gauss equation \cite[Cor. 2.7]{DL} on each $\Sigma_t$, the compactness of $V$ implies that the scalar curvature of the $\Sigma_t$ stays uniformly bounded.
Therefore, by standard results (e.g. the "Compactness Theorem" in Sect. 2 of \cite{Anderson}),  the foliation has a limit surface $\Sigma_\epsilon$, which is embedded, compact, and minimal. 
Moreover $\Sigma_\epsilon$ is two-sided: if not, then $\Sigma$  would have separated $V$. Therefore, by applying Lemma \ref{pen} again, there is a totally geodesic foliation near $\Sigma_\epsilon$. The maximum principle implies that $\Sigma_\epsilon$ is diffeomorphic to $\Sigma$.  A continuity argument now shows that $W \approx [0,1] \times \Sigma$ and each $\Sigma_t$ is totally geodesic. Thus $V$ is a surface bundle over $\mathbb{S}^1$ with totally geodesic fibers.   

To prove Theorem \ref{sym}, we proceed as above but invoke Lemma \ref{Kil} instead of Lemma \ref{pen}. (To this end, note that the minimal surface constructed above is least area.) If (i) of Lemma \ref{Kil} holds, then we proceed analogously as in the previous paragraph. If (ii) holds, then we obtain (iv) of Theorem 2 immediately. 

$\hfill \Box$ ~\\

\noindent {\bf Proof of Proposition \ref{haken}.}
Assume $M$ is past null geodesically complete.
The main result in \cite{Schoen} and Theorem 5.1 in \cite{Scott} yield
an incompressible, two-sided, least area immersion $f\colon\Sigma \to V$ of genus $g \ge 1$. (Here two-sided means that there is a global normal vector field defined on $f$.) This immersion is either embedded or double covers an embedded one-sided surface $K$. We discuss these cases in turn. 

\begin{description}
\item [I. $f$ is an incompressible, two-sided least area embedding.] By general covering space theory, there is a covering $p\colon \tilde{V} \to V$ such that $\pi_1(\tilde{V}) \cong \pi_1(\Sigma)$. By the lifting criterion, $f$ lifts to $F\colon\Sigma \to \tilde{V}$ which is a two-sided minimal embedding (in fact it is least area in its homotopy class by Theorem 3.4 in \cite{Scott}). 
   It must be that $\Sigma$ separates $\tilde{V}$. If not, then there is a loop in $\tilde{V}$ which traverses $\Sigma$ once and therefore generates an element in $\pi_1(\tilde{V}) \setminus \pi_1(\Sigma)$, contradicting the construction of $\tilde{V}$\footnote{This is a consequence of a more general fact. In short: If $\Sigma$ were nonseparating in $\tilde{V}$, then $\pi_1(\tilde{V})\cong A*_{\pi_1(\Sigma)}$ (see e.g. \cite[Proposition 1.2]{ScottNotes}) but then $\lvert\pi_1(\tilde{V}): \pi_1(\Sigma)\rvert = \infty$, a contradiction to our assumptions.}.    
    
    Since $\Sigma$ separates $\tilde{V}$, we have that $\tilde{V}\setminus \Sigma = U \sqcup W$ with $\partial U = \partial W = \Sigma$.
    Applying the Seifert-van Kampen theorem, we get the following commutative diagram
    {\small 
    \\\begin{tikzcd}
                                    &  & \pi_1(\bar{U}) \arrow[d, "j_1"] \arrow[rrd, "\phi_1"] &  &                                                                                  \\
    \pi_1(\Sigma) \arrow[rr] \arrow[rr, "i_*"] \arrow[rru, "i_1"] \arrow[rrd, "i_2"] &  & \pi_1(\tilde{V})                                &  & {\pi_1(\bar{U})*\pi_1(\bar{W})/\overline{\{ i_1(g)i_2(g)^{-1}, g \in \pi_1(\Sigma)\}}} \arrow[ll, "k"'] \\
                                                                                  &  & \pi_1(\bar{W}) \arrow[u, "j_2"] \arrow[rru, "\phi_2"] &  &
    \end{tikzcd}}

    By construction and Seifert-van Kampen $i_*$ and $k$ are isomorphisms.
    But then $i_1$ and $i_2$ are injective and  $j_1$ and $j_2$ are surjective. This means in particular that $\Si$ is incompressible 
    in $\bar{U}$, $\bar{W}$ and those manifolds are irreducible (see also \cite{Waldhausen}).
    From the injectivity of $i_1$ and $i_2$ we can conclude that $\phi_1$
    and $\phi_2$ are injective (Thm.\ 11.67 in \cite{Rotman}) and thus  
    $j_1$ and $j_2$ must be isomorphisms. (See 17.2 in \cite{Hall}.)
   
Aiming at a contradiction, assume $\bar{U}$ is compact. Then  $\pi_1(\bar{U}) \cong \pi_1(\Sigma)$ together with Thm.\ 10.2 in \cite{Hempel} implies  that $\bar{U}$ is a trivial $I$-bundle over $\Sigma$. This entails that $\bar{U}$ has two disjoint boundary components. But from above, $\Sigma$ is separating and $\tilde{V}$ has no boundary, which is a contradiction. It follows that $\bar{U}$ is noncompact, and the same reasoning applies to $\bar{W}$ as well.
    
    With these arguments at hand we are now in a situation where we can apply Lemma \ref{pen}. Thus there exists a neighborhood $B$ of $\Sigma$ in $V$ such that $B$ is diffeomorphic to $(-\epsilon, \epsilon) \times \Sigma$ and $\Sigma_t = \{t\} \times \Sigma$ is totally geodesic within $V$ and a MOTS within $M$.
    As in the previous case, we can apply a compactness theorem to obtain an embedded, compact, totally geodesic limit surface $\Sigma_\epsilon$. 
    
    If $\Sigma_\epsilon$ is two-sided, we can apply Lemma \ref{pen} to again obtain a local foliation $\Sigma_{\epsilon} \times (-\delta, \delta)$ for some $\delta > 0$.  But this foliation intersects $B$ and thus by the maximum principle for minimal surfaces, we have $\Sigma_\epsilon \cong \Sigma$. This extends $B$ to $B_\delta \cong \Sigma \times (-\epsilon, \epsilon + \delta)$. We can apply the compactness theorem to obtain a limit surface $\Sigma_{\epsilon + \delta}$. Aiming at a contradiction, assume all limit surfaces are two-sided, we can repeat these arguments indefinitely. By an open and closed argument we get a contradiction to the compactness of $V$.

    
    Thus, there are nonorientable limit surfaces $\Sigma_1$ and $\Sigma_2$.  By gluing two copies of $V\setminus (\Sigma_1 \cup \Sigma_2)$ along their two boundary components, we get a double cover $V'$ of $V$. The boundary components of $V\setminus (\Sigma_1 \cup \Sigma_2)$ are diffeomorphic to $\Sigma$ (this can be seen via \cite[Thm.\ 10.5]{Hempel} or another maximum principle argument within $V'$). Thus, $V'$ is a surface bundle over $\mathbb{S}^1$ with fiber $\Sigma$. Hence $\Sigma_1$, $\Sigma_2$ are double covered by $\Sigma$. Subsequently, $\Sigma_1\cong\Sigma_2$ by the classification theorem of closed surfaces. Therefore, by construction, $V$ has a semibundle structure, meaning it is the union of two twisted $I$-bundles over $\Sigma_1\cong\Sigma_2$ glued along their common boundary $\Sigma$. That is, $V = V_1 \cup V_2$ where $V_i \cong \Sigma_i \tilde{\times} I$. 

 \item[II. $f$ double covers a one-sided surface $K$]. We can go to the double cover $\tilde{V}$ such that the lift $\tilde{K}$ of $K$ is a connected, orientable, separating, incompressible embedded minimal surface. Thus, by our previous arguments used in I, $\tilde{V}$ has a semibundle structure. Furthermore by construction both components of $\tilde{V}\setminus \tilde{K}$ are diffeomorphic to $V \setminus K$. But then $V$ is diffeomorphic to $(\tilde{K} \tilde{\times} I \setminus \tilde{K}) \cup K$, i.e. it itself has a semibundle structure (in particular $\tilde{K}$ is totally geodesic in $V$).

\end{description}
$\hfill \Box$ ~\\

\noindent {\bf Proof of  Proposition \ref{orient}.} 
Assume $M$ is past null geodesically complete. There are three separate cases to consider:
\begin{description}
        \item[I. $V$ is reducible.] We first show that $V$ has to be prime. Let $\tilde{V}$ be its orientable double cover. Apply Theorem \ref{new} to the corresponding covering spacetime $\tilde{M}$. Therefore $\tilde{V}$ is finitely covered by a surface bundle over $\mathbb{S}^1$ -- call it $\tilde{V}'$. We see that $\tilde{V}'$ is prime: If the genus of the fibers is $\geq 1$, then its universal cover is $\mathbb{R}^3$ and hence irreducible. This contradicts the assumed reducibility of $V$, \cite[Prop. 1.6]{Hatcher3}. If the genus is zero, then $\tilde{V}'$ is homeomorphic to $\mathbb{S}^1 \times \mathbb{S}^2$ and hence prime.  
        
        To finish the argument, we use the following fact \cite[p.\ 7]{Hatcher}, which we prove below. Fact: The only non-prime manifold covered by a prime manifold is $\mathbb{RP}^3\#\mathbb{RP}^3$ double covered by $\mathbb{S}^1 \times \mathbb{S}^2$. 

        Since $\mathbb{RP}^3 \#\mathbb{RP}^3$ is orientable and $V$ is not, the fact implies that $V$ is prime. Therefore $V$ is homeomorphic to  $\mathbb{S}^1\tilde{\times}\mathbb{S}^2$ by reducibility. Then $\tilde{V}$ is homeomrophic to $\mathbb{S}^1 \times \mathbb{S}^2$. Theorem \ref{new} implies that $\tilde{V}$ is foliated by totally geodesic two-spheres. This foliation maps to a foliation of totally geodesic two-spheres within $V$. 

        Proof of fact: Suppose $V = V_1 \# V_2$ is not prime and it's covered by an orientable prime manifold -- call it $\tilde{V}$. Since $\tilde{V}$ is reducible, it follows that $\tilde{V}$ is homeomorphic to $\mathbb{S}^1 \times \mathbb{S}^2$ or $\mathbb{S}^1 \tilde{\times} \mathbb{S}^2$. In either case, $\mathbb{Z}$ is a subgroup of $\pi_1(V)$ with finite index, so by Lemma 11.4 in \cite{Hempel}, $\pi_1(V)$ contains a finite normal subgroup $K$ with $\pi_1(V)/K$ isomorphic to either $\mathbb{Z}$ or $\mathbb{Z}_2*\mathbb{Z}_2$. $K$ must be the trivial subgroup, otherwise it contradicts Lemma 11.2 in \cite{Hempel}. Thus $\pi_1(V) = \mathbb{Z}_2*\mathbb{Z}_2$. Then by Thm.\ 7.1 in \cite{Hempel}, along with the positive resolution of the elliptization conjecture, it follows that $V_1$ and $V_2$ are homeomorphic to $\mathbb{RP}^3$.

        \item [II. $V$ is irreducible and $\pi_2(V) \neq 0$.] Then $V$ contains an embedded 2-sided projective plane $\mathbb{RP}^2$ by the projective plane theorem (see \cite[Thm. 1.1]{Epstein} or \cite[Thm. 4.12]{Hempel}).
        It follows that there exists an embedded stable least area surface $\Sigma$ homeomorphic to $\mathbb{RP}^2$  \cite[Prop. 2.3]{Bray}. Moreover, $\Sigma$ is necessarily 2-sided \cite[Lem. 2]{Heil}.
        Note that $0 \neq [\Sigma]  \in \pi_2(V)$ since $\Sigma$ is incompressible in $V$ by Proposition 2.1 in \cite{Bray}.
        By Lemma 2.1 in \cite{Swarup} the fundamental group of the orientable double cover $\tilde{V}$ of $V$ is torsion-free. Since $\tilde{V}$ is reducible (by the sphere theorem), part I of this proof shows that $\tilde{V}$ is either $\mathbb{S}^1 \times \mathbb{S}^2$ or $\mathbb{RP}^3\# \mathbb{RP}^3$ but only the former has torsion-free fundamental group.  Furthermore \cite[Lem. 2.1]{Swarup} also implies that $V$ is homotopic to $V' := \mathbb{S}^1\times \mathbb{RP}^2$.

        Since $\pi_2(V') \cong \pi_2(\mathbb{RP}^2)$ there is a homotopy from $\Sigma$ in $V$ to a nonseparating embedded fiber $\Sigma' \cong \mathbb{RP}^2$ in $V'$. Therefore there is a non-trivial loop in $c'$ in $V'$ which intersects $\Sigma'$ transversally only once. We remark that transversality is a generic condition and thus by the homotopy from $V$ to $V'$, this loop is  homotopic to a non-trivial loop $c$ in $V$ which intersects $\Sigma$ transversally only once (mod 2) (see also Chapters 2.3 and 2.4 in \cite{Pollack}, in particular exercise 2 in section 2.4). Thus $\Sigma$ is nonseparating.
        
        Therefore we can pass to a noncompact cover and use arguments\footnote{Note that the construction of the cover presented in the proof of Theorem \ref{new} only used that $\Sigma$ was nonseparating and two-sided; orientability was not used.} as in Theorem \ref{new} to deduce that $V$ is a surface bundle over $\mathbb{S}^1$ with totally geodesic fibers diffeomorphic to $\mathbb{RP}^2$.   
        

        \item[III. $V$ is irreducible and $\pi_2(V) = 0$.]

         We claim and prove in the next paragraph that there is a two-sided least area incompressible nonseparating embedding of $\Sigma$ into $V$ (with $\Sigma \neq \mathbb{S}^2, \mathbb{RP}^2$).  Then we can  apply the arguments of the proof of Theorem \ref{new} to conclude that $V$ is a surface bundle over a circle. 
        
        We now prove the claim. First, since $V$ is nonorientable, by \cite[Lemmas 6.6 and 6.7]{Hempel} it contains a two-sided, incompressible, nonseparating embedding $g \colon \Sigma \to V$. Moreover, $V$ contains no two-sided $\mathbb{RP}^2$ since $\pi_2(V) = 0$; therefore it's $P^2$-irreducible and hence a Haken manifold.  There exists a two-sided incompressible least area immersion $f\colon\Sigma \to V$ homotopic to $g$ \cite[Thm. 3.1]{Scott}. Then either $f$ is an embedding or it double covers a one-sided surface $K$ and $g(\Sigma)$ bounds a submanifold of $V$ which is a twisted $I$-bundle over a surface isotopic to $K$ \cite[Thm. 5.1]{Scott}. However, the latter case contradicts the fact that $g$ is nonseparating. Lastly, since $g$ is nonseparating and $f$ is homotopic to $g$, it follows that $f$ is also nonseparating since the intersection number mod 2 is preserved under homotopy (see the corollary on page 79 of \cite{Pollack}).

        \end{description}
$\hfill  \Box$ ~\\

We give examples where (ii) of Proposition \ref{orient} applies for each of the three scenarios  in  the proof. For I, consider the Nariai spacetime (Example 4 in Section \ref{examples}) quotiented by a twist on the spatial slices yielding the orientatable double covering $\mathbb{S}^1 \times \mathbb{S}^2 \to \mathbb{S}^1 \tilde{\times} \mathbb{S}^2$. Similarly, for II, quotient the Nariai spacetime via its spatial slices $\mathbb{S}^2 \times \mathbb{S}^1 \to \mathbb{S}^1 \times \mathbb{RP}^2$. For III, consider a flat vacuum spacetime (Example 8 in Section \ref{examples}) with spatial topology $\mathbb{S}^1 \times \mathbb K$ where $\mathbb K$ is the Klein bottle. 


\medskip

\noindent {\bf Proof of Proposition \ref{nonprime}.} Assume that $M$ is past null geodesically complete. From the first part of the proof of Proposition \ref{orient}, we know that $V$ is homeomorphic to $\mathbb{RP}^3 \#\mathbb{RP}^3$.


This implies that there is an embedding $g\colon \mathbb{S}^2 \to V$ which separates and generates a non-trivial element of $\pi_2(V)$ \cite[Prop. 3.7 and 3.10]{Hatcher3}. Then, by \cite[Thm. 7]{Meeks} and \cite[Thm. 5.8]{Sacks}, $g$ is homotopic to a non-trivial minimal immersion $f \colon \mathbb{S}^2 \to V$ which is either embedded or double covers an embedded projective plane. 
In the former case we can apply the same arguments as in the previous results to obtain a totally geodesic fibration within $V$  where the fibers are MOTS/MITS within $M$.
    In the latter case we can lift $f$ to a double cover $V'$, which is also homeomorphic to $\mathbb{RP}^3\#\mathbb{RP}^3$, to find a minimal embedding $f'\colon \mathbb{S}^2 \to V$ homotopic to $f$. 

$\hfill \Box$

\begin{remark}
    The proof of Theorems \ref{new} and \ref{Kil} relied on the positive resolution of the "virtual $b_1>0$ conjecture" to obtain a minimal  genus $g \geq 1$ surface $ \Sigma$ embedded within a noncompact cover $\tilde{V}$ of $V$. Although not needed for this paper, we remark another way in which one can obtain a minimal embedding. If $V$ is a 3-dimensional closed, irreducible, orientable manifold with $|\pi_1(V)| = \infty$, then the positive resolution of the surface subgroup conjecture \cite{Kahn} along with results from \cite{Schoen} shows that there is a genus $g \geq 1$ surface $\Sigma$ and a least area immersion $f \colon \Sigma \to V$. Consider the noncompact covering $p \colon \tilde{V} \to V$ as in the proof Proposition \ref{haken}  and the lift $F \colon \Sigma \to \tilde{V}$. Since $V$ is irreducible, so is $\tilde{V}$ ($\pi_2(\tilde{V}) = 0$ by the homotopy lifting property of coverings). Therefore $F$ is a homotopy equivalence. Then, by Thm.\ 2.1 in \cite{Scott}, it follows that $F \colon \Sigma \to \tilde{V}$ is an embedding.

\end{remark}
    

\section{Discussion, Examples}

\label{examples}


While Theorems \ref{new} and \ref{sym} cover a wide range of manifolds, their motivation stems from General Relativity, i.e. from solutions of Einstein's equations. In the following examples we restrict ourselves to the vacuum case, i.e. $\mathrm{Ric} = \Lambda g$, $\Lambda \ge 0$; in fact we take $\Lambda > 0$ except for Examples \ref{ex2}, \ref{ex8} and \ref{ex9}.

The topology of $V$ is spherical in Examples \ref{ex1} and \ref{ex2}, and $\mathbb{S}^2 \times \mathbb{S}^1$ in Examples \ref{ex3}--\ref{ex7}. While the $\mathbb{S}^2$-factors are round in Examples \ref{ex3}--\ref{ex5}, they are only axially symmetric in Examples \ref{ex6} and \ref{ex7}, but with  additional ($t$-$\phi$)-symmetries in the data, as defined below. As to the $\mathbb{S}^1$-direction, Examples \ref{ex3}, \ref{ex4} and \ref{ex6} enjoy the corresponding continuous symmetry. In fact these three examples have symmetry groups containing $U(1) \times U(1)$ and satisfy (\ref{Kdec}) with respect to both Killing fields. They thus provide examples for the simultaneous appearance of cases (iii) and (iv) in Theorem \ref{sym}, cf.\ Remark \ref{amb}.  The corresponding spacetimes $M$ have trivial (Example \ref{ex3}) rather subtle (Example \ref{ex4}) or largely unknown (in)completeness structures (Example \ref{ex6}, Remark \ref{rem7}).  
On the other hand, in Examples \ref{ex5} and \ref{ex7} there is no symmetry in the $\mathbb{S}^1$-direction, and the area of the $\mathbb{S}^2$-sections has minima and maxima (i.e.\ strictly stable or strictly unstable extrema). This entails horizons and incomplete geodesics according to cases (i) in Theorems \ref{new} and \ref{sym}. 
Finally we consider locally flat and hyperbolic $K(\pi,1)$ geometries in Examples \ref{ex8} and \ref{ex9}.

Our Theorems \ref{new} and \ref{sym} are admittedly of limited value for predicting the 
singularity structure of a spacetime from initial data. However, in Examples \ref{ex4} and \ref{ex9} geodesic incompleteness was (or would be) non-trivial to obtain by other means. In any case, the value of our results rather lies in providing a classification of the general case, i.e.\ without symmetries and beyond solutions of Einstein's equations. We nevertheless believe that all subsequent examples are useful to illustrate the applicability of our results and some methods of the proofs.

\begin{example}[Spherical slices in  de Sitter spacetime]\label{ex1}~\\
We consider de Sitter spacetime given by the manifold $\mathbb{R} \times \mathbb{S}^3$ and metric given by 
\begin{eqnarray}
ds^2 & = & -dt^2 + \frac{3}{\Lambda} \cosh^2 \left(\sqrt{\frac{\Lambda}{3}} t\right) [d\tau^2 + \sin^2 \tau (d\theta^2 + \sin^2
\theta  d\phi^2)]  \nonumber \\
&  &   t \in (-\infty, \infty) \quad \tau, \theta \in [0, \pi) \quad \phi \in [0,2\pi).
\end{eqnarray}
The extrinsic curvature of the round spheres $\mathbb{S}^3$ at $t= const.$ reads
\begin{equation}
{\K} =  \frac{1}{2} \frac{d}{dt} g = \sqrt{\frac{\Lambda}{3}} \tanh \left(\sqrt{\frac{\Lambda}{3}}t\right)g.
\end{equation}
Theorem 0 applies to all round spheres 
$\mathbb{S}^3$ for all $t> 0.$ (or quotients thereof), while Theorem 1 applies for all $t\geq 0$. The well-known geodesic completeness is consistent with the spherical topology of the Cauchy surfaces.
\end{example}

\begin{example}[The Taub-NUT spacetime \cite{Taub} \cite{NUT}]\label{ex2}~\\
This 2-parameter family of $\Lambda = 0$ solutions can be written as
\begin{equation}
ds^2 = - U^{-1} dt^2  + 4\ell^2 U (d \psi + \cos \ta d\phi)^2 + (t^2 + \ell^2) (d \ta^2 + \sin^2 \ta d\phi^2)
\nonumber
\end{equation}
where
$U(t) = -1 + \frac{2(mt + \ell^2)}{t^2 + \ell^2}$ and $m, \ell \in \mathbb{R}$. 

The  $t = \mathrm{const.}$ surfaces are diffeomorphic to $\mathbb{S}^3$, which is parametrized by the Euler angles $ \psi \in [0, 4\pi);~\ta \in [0, \pi), ~\phi \in [0, 2\pi)$.
The "Taub-region" $U > 0 $ is given by $m - \sqrt{m^2 + \ell^2} < t <  m +\sqrt{m^2 + \ell^2}. $

Selecting now ($V,g_{ij})$ to be the surface $t = 0$, the extrinsic curvature reads\\
\[  {\K}  =  \frac{\sqrt{U}}{2} \frac{\partial}{\partial t} g=
4m \left( \begin{array}{l} 1 \quad~~~~ 0 \quad \cos\ta \\
                        0 \quad ~~~~0 \quad ~~0 \\
                             \cos \ta ~~~  0 \quad \cos^2 \ta  \end{array} \right)  \cong 
                     4m \left( \begin{array}{l} 1 + \cos^2 \ta \quad 0 \quad\quad 0 \\
                       \quad ~~~  0 \quad~~~~~ 0 ~~~ \quad 0 \\
                   ~~~ \quad \, 0 \quad \quad ~ 0 \quad \quad 0   \end{array} \right),
\]
where $\cong$ denotes diagonalisation. This matrix is obviously $2$-convex which allows application of Theorem \ref{new}. While the Taub region is known to be null geodesically incomplete (as a maximally extended globally hyperbolic spacetime), $V$ is obviously a spherical space, cf.\ Remark \ref{rem1}. 
\end{example}

\begin{example}[$\mathbb{S}^2 \times \mathbb{S}^1$ slices in Nariai spacetime]\label{ex3}~\\ 
The Nariai spacetime is  a geodesically complete solution which reads
\begin{eqnarray}
\label{Nar}
ds^2  &= & -dt^2 + \frac{1}{\Lambda} \left( \cosh^2 \sqrt{\Lambda} t\, d\al^2 + d\theta^2 + \sin^2 \theta  d\phi^2\right)   \\
&  &   t \in (-\infty, \infty)  \quad  \al  \in [0, L), ~ L \in \mathbb{R}, \quad \theta \in [0,\pi) \quad \phi \in [0,2\pi). \nonumber
\end{eqnarray}

The $t = const.$ slices have topology $\mathbb{S}^2 \times \mathbb{S}^1$; we call the $t=0$ slice the "Nariai torus" henceforth. 
We obtain for the only non-vanishing  component of the extrinsic curvature $\K$
\begin{equation}
{\K}_{\al\al} =  \frac{1}{2} \frac{d}{dt} g_{\al\al} = {\Lambda}^{-1/2} \cosh \sqrt{\Lambda}t ~ \sinh \sqrt{\Lambda}t.
\end{equation}
Therefore, $\K$ is positive semidefinite for all $t \ge 0$.  Now Theorem \ref{new} applies with consequence (iii), while Theorem \ref{sym} also applies, with conclusion (iii) with respect to the Killing vector $\partial/\partial \phi$ but conclusion (iv) with respect to $\partial/ \partial \alpha$. In this case, we need not pass to a finite cover to obtain the rigidity results.
\end{example}

\begin{example}[Products of constant curvature  metrics]\label{ex4} ~\\
Fajman and Kr\"oncke \cite{FK} have analyzed  spacetimes where the spatial sections are products of Einstein spaces. In $3$ spatial dimensions their ansatz takes the form \cite[Thm. 1.6]{FK} 
\begin{equation}
\label{mFK}
ds^2 =  -dt^2 + \frac{b(t)^2}{\Lambda} d\alpha^2  + \frac{a(t)^2}{\widehat \Lambda} 
[d\theta^2 + \sin^2\theta  d\phi^2] 
\end{equation} 
where $\widehat \Lambda$ is a constant.
The authors consider the initial value problem  with $a(0) = 1$ and $[db/dt] (0) = 0$.
Then the constraint implies $\widehat \Lambda \le \Lambda$.  While 
 $\widehat \Lambda = \Lambda$ reduces to the Nariai case $a \equiv 1$,
 $\widehat \Lambda  < \Lambda$, yields incompleteness in the sense of diverging curvature 
 scalars. 
 
 As to our results, at $t=0$ the extrinsic curvature

\begin{equation}
{\K} =  \frac{1}{2 } \frac{d}{dt} g =  {\widehat \Lambda}^{-1}\,\frac{da}{dt}\,
\mbox{diag}[0, 1,  \sin^2\theta] 
\end{equation}
is strictly $2$-convex. (The degeneracy at $\theta = 0$ is of course an artifact of the polar coordinates.) Since this $t=0$ section of (\ref{mFK}) is obviously not a spherical space, Theorem 0 yields null geodesic incompleteness which is consistent with the findings of Thm.\ 1.6.\ in \cite{FK}. In this example Theorems \ref{new} and  \ref{sym} are inconclusive as (\ref{mFK}) is also consistent with consequences (iii) and (iv).
\end{example}

\begin{example}[Time-symmetric slices in Kottler spacetime]\label{ex5} ~ \\
In the well-known Kottler (also Schwarzschild de Sitter) spacetime we restrict ourselves to the time-symmetric slices. They are conformal to  the Nariai torus with  metric 
\begin {equation}
ds^2 = r^2 \left( \frac{dr^2}{\sigma} + d\theta^2 + \sin^2 \theta d\phi^2\right) = r(\al, m ,\Lambda)^2 \left( d\al^2 + d\theta^2 + \sin^2 \theta d\phi^2\right).
\end{equation}
Here $m \in \mathbb{R}$, $m < 1/(3\sqrt{\Lambda})$, $\sigma = (r^2 -2mr - \Lambda r^4/3)$. 

The positive zeros 
 $r = r_b$ and $r = r_c$  of $\sigma$ give the black hole and the cosmological horizons. The coordinates are obviously related via
\begin{equation}
\label{alr}
\frac{d \al}{dr} = \sigma^{-\frac{1}{2}} \qquad r(\al = 0) = r_b.
\end{equation}
The following equation in which we introduce the length $L$ of the Nariai torus and the period $P(m,\Lambda)$ 
 states that the period has to "fit"  on the torus, i.e.

\begin{equation}
\label{per1}
L = \int_0^L d\alpha  = 2 j \int_{r_b}^{r_c} \sigma^{-\frac{1}{2}} dr =  j P(m,\Lambda) 
\end{equation}
for some $j \in \mathbb{N}$.  Thus the Cauchy surface contains $j$ pairs of black hole and cosmological horizons, and (\ref{per1}) entails a relation between the parameters $m$, $L$ and $j$. 
As is well-known and suggested by the "horizon" terminology, the spacetime is (future and past) null geodesically incomplete. Thus we are in case (i) of Theorems \ref{new} and \ref{sym}; the other cases are ruled out since $\partial/\partial \alpha$ is not Killing and the foliation is not totally geodesic.

\end{example}

In connection with the following two examples, we revisit the notion of ($t$-$\phi$)-symmetric data. 
For data with Killing field $\xi$ we first recall some definitions from Section \ref{int} (before Thm \ref{sym}):    $\|\xi\|$ is the norm of $\xi$, $\zeta = \|\xi\|^{-1} \xi$ its normalisation and  the 
$(1,1)-$ tensor ${\cal P} = {\cal I} -  (\zeta \otimes \zeta^*)$ projecting onto the 2-space orthogonal to $\xi$,  where ${\cal I}$ is the identity. We denote the components of $\K$ with respect to this decomposition 
  by ${\K} (\zeta,\zeta)$, ${\K}(\zeta, \bot) = \zeta^* {\K}  {\cal P} $ and ${\cal K}^{\bot}  = {\K(\bot,\bot)}  = {\cal P}^T {\K}{\cal P} $. 
  
The following definition seems to appear first in \cite{Bardeen} (eqs. (D5) and (D6)). There (and elsewhere) 
it is used  in connection with axially symmetric data, (i.e. the isometry has fixed points) which we do not require.

\begin{defn}["($t$-$\phi$)-symmetric" data] ~\\ 
Initial data $(V,g,{\K})$ are called "($t$-$\phi$)-symmetric"  if they are $U(1)$ symmetric with Killing 
field $\xi$ such that ${\K} (\xi,\xi)$  and $\K (\bot, \bot)$  vanish identically.
\end{defn}

This definition captures the idea that the spacetime should be invariant under the simultaneous change of the direction of time and rotation: Such changes reverse the signs of $\K$ and $\xi$, respectively,  and invariance requires that all  products containing an odd number of these tensors  have to vanish. We are thus left with $\kappa = \K (\zeta, \bot)$ as the only non-vanishing components. We further observe that  the eigenvalues of $\K$ are $\{-c,0,c\}$ where $c$ is the norm of the 2-vector $\kappa$. In particular, the
data are maximal.
Now this class is obviously admitted by condition (\ref{Kdec}) and therefore covered by Theorem \ref{sym}, but incompatible with $2$-convexity as required in Theorem \ref{new} unless $\K$ vanishes identically.

\begin{example}["Squashed Nariai" data]\label{ex6}~\\
As in Example \ref{ex5} we consider data for Einstein's equations with $\La > 0$ conformal to the Nariai torus but now with conformal factor  depending on $\theta$ rather than $\alpha$, viz. 
\begin {equation}
\label{sn}
ds^2 =   \varphi(\theta, J, \Lambda)^4 \left( d\al^2 + d\vt^2 + \sin^2 \vt d\phi^2\right).
\end{equation}

Here the constant $J$
is called the angular momentum---as the name suggests the data contain a corresponding "momentum"  ${\K}(\vt, J,\La)$. We sketch the proof that such $\K$ and $\vp$ exist, which is rather intricate, cf.\ \cite{BPS,SP} for details.

We first recall the (suitably adapted) conformal method for solving the constraints. 
We assume we are given an arbitrary  "seed manifold"  $(\wh V,\wh g)$ together with  a trace-free, divergence-free  tensor $\widehat {\K}$. We look for initial data solving the constraints
\begin{equation}
\label{con}
\mbox{div}_g\,  {\K} = 0  \qquad {}^gR = 2 \Lambda +  {\cal C}_{g}( \K \otimes \K)  
\end{equation}
where ${}^g R$ is the scalar curvature of $(V,g)$ and ${\cal C}_g$ is a double contraction. 
From the conformal behavior of the scalar curvature it is clear that
\begin{equation}
{\K} = \vp^{-2} \wh{\K }    \qquad g = \vp^{4} \wh g
\end{equation}
 solve  (\ref{con}) provided $\vp$ satisfies the "Lichnerowicz"-equation  on $(\wh V, \wh g)$,
\begin{equation}
\label{Lich}
\left(\wh \Delta - \frac{1}{8} \wh R \right) \vp = 
- \frac{1}{4} \Lambda \vp^5 - \frac{ {\cal C}_{\wh g}(\wh {\K} \otimes \wh{\K})  }{8 \vp^7}\,.
\end{equation}
We now choose for $(\wh V,\wh g)$ the Nariai torus and for $\wh {\K}$  the symmetric, trace-free tensor
\begin{equation}
\label{Ksn}
\wh {\K}  =  3J\Lambda^{5/2}\, \aleph \otimes_s \Phi
\end{equation}
where $\aleph = \partial/\partial \alpha$ and $\Phi = \partial/\partial \phi$ are the orthogonal Killing vectors on the torus, and $\otimes_s$ is the symmetrized tensor product. This tensor is indeed divergence-free, i.e. $ {\mbox{div}}_{\wh g}\,\wh {\K}  =0$. 
 
It remains to solve (\ref{Lich}) with the input from above. From results by Premoselli
\cite{BP}, there is a  $J_{\max} \in \mathbb{R}^+_0$ such that for each $J < J_{\max}$, there exists a unique smooth, positive solution $\vp_J$ of  (\ref{Lich}) which is stricty  stable (in the sense that the lowest eigenvalue of the linearisation of (\ref{Lich}) at $\vp_J$ is positive). Moreover, for $J = J_{\max}$ there is a unique marginally stable solution (i.e.\ the  lowest eigenvalue vanishes).

Each of these solutions $\vp_J$ now leads to a solution of the constraints $(V, g, {\K})$ 
which (due to their stability)  share the $U(1) \times U(1)$ symmetry of the coefficients in the 
Lichnerowicz equation (\ref {Lich}), cf.\ e.g.\ \cite[Prop.\ 2]{BPS}. This implies that  the metric indeed takes the form (\ref{sn}), which is a totally geodesic foliation.  Moreover, it is obvious from (\ref{Ksn}) that the data are both ($t$-$\phi$)-symmetric as well as $(t-\alpha)$ symmetric 
as defined above. The constant $J$ is the  "Komar" angular momentum of the data with respect 
to the Killing field $\partial/\partial \phi$.

We now observe that our Theorem \ref{sym} does apply (w.r.t.\ both Killing vectors $\Phi$ and $\aleph$) 
but does not allow to draw any conclusions on (in)completeness of the evolved spacetime.
 \end{example}

\begin{remark} 
\label{rem7}
More generally, there has  been studied the case that a compact and connected two-dimensional Lie group acts effectively on $V$ (cf. \cite{HR} for a review). 
 This reduces to the so-called Gowdy class in the case  of vanishing "twist constants", and
 the latter vanish automatically for $\Lambda$-vacuum data and topology  $\mathbb{S}^2 \times \mathbb{S}^1$.  Geodesic (in)completness  for this class of data has been studied in \cite{Beyer_Nariai, BEF},
 mainly by numerical methods. It would be interesting to investigate the applicability of our Theorem \ref{sym} in this context.
\end{remark}

\begin{example}[Kerr-de Sitter]\label{ex7} ~\\
This is a well-known two-parameter family of stationary, axially symmetric  black hole spacetimes. 
In Boyer-Lindquist coordinates the induced metric on maximal slices reads 
\begin {equation}
\label{KdS}
ds^2 = \rho^2 \left( \frac{dr^2}{\sigma} + \frac{d\theta^2}{\chi}\right) + 
\frac{\sin^2 \theta}{\kappa^2 \rho^2}\left[\chi (r^2 + a^2)^2 - \sigma a^2 \sin^2  \theta  \right] d\phi^2,
\end{equation}
where $m$, $a$ and $\kappa = 1 + \Lambda a^2/3$ are constants and
\begin{equation}
\label{src}
\sigma = (r^2 + a^2) \left( 1 - \frac{\Lambda r^2}{3} \right) - 2mr, \quad \rho^2 = r^2 + a^2 \cos^2 \theta,  \quad   \chi = 1 + \frac{\Lambda a^2 \cos^2 \theta}{3}\,.                   \end{equation}
The constants $m$ and $a$ satisfy bounds given by the extreme solutions, but also "absolute" bounds in terms of $\Lambda$ alone, cf.\ e.g.\ \cite{CCK,Oelz2013} for details. The data are  $(t,\phi)$-symmetric.
As in the Kottler case of Example \ref{ex5} we restrict ourselves 
to the region $r \in [r_b, r_c]$ outside the horizons  where  $\sigma  \ge 0$. At $\sigma = 0$ the metric  (\ref{KdS}) is regularized by replacing $r$ by $\al$ defined via (\ref{alr}) but with $\sigma$ from  (\ref{src}).
While the solution is no longer conformal to the Nariai torus,  we can still fit the metric on a  $\mathbb{S}^2 \times \mathbb{S}^1$ manifold of length $L$ provided the period $P(m,a,\Lambda)$ 
satisfies

\begin{equation}
\label{per}
L = \int_0^L d\alpha  = 2 j \int_{r_b}^{r_c} \sigma^{-\frac{1}{2}} dr =  j P(m,a,\Lambda) \qquad j \in \mathbb{N}.
\end{equation}

and we still obtain an array of $j$ pairs of  horizons \cite{JC}.
In consistency with well-known facts and in analogy with Example \ref{ex5}, our  Theorem \ref{sym} yields  null geodesic incompleteness via point (i).
\end{example}

\begin{example}[Flat vacuum initial data sets]\label{ex8}~\\
Let $V$ be a closed flat three-dimensional manifold equipped with time-symmetric vacuum initial data for the Einstein equations. The corresponding maximal globally hyperbolic vacuum development is the Lorentzian product $\mathbb{R} \times V$, which is clearly geodesically complete. There are 10 such data sets of the form $V = \mathbb E^3/\Gamma$  where  $\Gamma$ is a discrete and fixed-point free symmetry group of the Euclidean space $\mathbb E^3$. There are six orientable ones $G_1, \dotsc, G_6$ and four nonorientable ones $B_1, \dotsc, B_4$ (see the classification in \cite{wolf}.)
For each one Theorem 0 does not apply, but Theorem 1 does. If $V$ is either the 
flat torus $G_1 = \mathbb T^3$ or one of other 4 possibilities with  $H_2(V,\mathbb{Z}) \neq  0$, 
then Theorem 1  (iii) applies directly to $V$ (without going to a cover).
On the other hand,  $G_6$ (i.e., the Hantzsche–Wendt manifold),  has  $H_2(G_6,\mathbb{Z}) = 0$ and is not a surface bundle over the circle. Now Theorem 1 (iii) still applies to a cover. Indeed, if one applies the construction in the proof of Theorem \ref{new} to $G_6$, then one obtains its double cover $G_2$. However, as mentioned in the introduction, Proposition \ref{haken} applies directly to $G_6$ (i.e. without invoking a cover). \end{example}

\begin{example}[Hyperbolic initial data sets]\label{ex9}~\\ 
Let $(V,g,{\K})$ be an initial data set for the vacuum Einstein equations with or without a cosmological constant. Assume $g$ is a hyperbolic metric on a closed 3-manifold $V$ and $\K$ is $2$-convex. Then Theorem \ref{new} implies that the resulting maximal globally hyperbolic development is past null geodesically incomplete. For if this was not the case, then there would be a three-manifold with a hyperbolic metric which is a surface bundle over $\mathbb{S}^1$ containing totally geodesic fibers. But this is impossible; see \cite[Thm. 3.3]{Uhlenbec} and the discussion on mapping tori below it. 
\end{example}

It would be interesting to know if there exists a Riemannian metric $h$ on a 3-manifold $V$ with hyperbolic structure such that $(V,h)$ embeds into a spacetime $(M,g)$ where Theorem \ref{new} (iii) applies. Example \ref{ex9} shows that $h$ cannot be taken to be the hyperbolic metric. 

\section*{Acknowledgments.}
We are grateful to  Greg Galloway for fruitful interaction in all stages of the present work, most importantly in connection with the virtual positive $b_1$-conjecture (cf.\ Remark \ref{Greg}).  We also acknowledge helpful correspondence with Florian Beyer,  Shuli Chen,  Klaus Kr\"oncke, Vanderson Lima and Franco Vargas Pallete. Furthermore, we thank the referee for useful comments.

Eric Ling was supported by Carlsberg Foundation CF21-0680 and Danmarks Grundforskningsfond CPH-GEOTOP-DNRF151. Carl Rossdeutscher and Walter Simon were funded by the Austrian Science Fund (FWF) [Grant DOI 10.55776/ \linebreak P35078]; 
Roland Steinbauer was  funded in part by FWF as well [Grant DOI 10.55776\linebreak/EFP6]. Part of the research on this paper was performed while two of the authors were participants of the program “New Frontiers in Curvature: Flows, General Relativity, Minimal Submanifolds, and Symmetry” at the Simons Laufer Mathematical Sciences Institute (SLMath). We thank SLMath for its support. We also acknowledge the support of the Simons Center for Geometry and Physics, where parts of this work were carried out during the program 
\emph{``Geometry and Convergence in Mathematical General Relativity''}.  For open access purposes, the authors have applied a CC BY public copyright license to any author accepted manuscript version arising from this submission.

\medskip

\bibliographystyle{unsrt}
\bibliography{references.bib}
\end{document}